\RequirePackage{fix-cm}
\documentclass[twocolumn,epjc3]{svjour3}  
\smartqed  
\RequirePackage{graphicx}
\usepackage{xcolor}
\usepackage{graphicx}
\usepackage{main-defs}
\usepackage{amsmath}
\def\makeheadbox{}

\begin{document}

\title{Machine Learning Enables Real-Time Waveform Decomposition for Dual-Readout Calorimetry}

\author{Liangyu Wu\thanksref{addr0,addr1,e1}
        \and
        Qibin Liu\thanksref{addr0}
        \and
        Marco Toliman Lucchini\thanksref{addr2,addr3}
        \and 
        Julia Gonski\thanksref{addr0}
        \and Marcello Campajola\thanksref{addr4,addr5}
        \and Stefano Moneta\thanksref{addr6}
}

\thankstext{e1}{e-mail: liangyu@slac.stanford.edu}


\institute{SLAC National Accelerator Laboratory, 2575 Sand Hill Rd, Menlo Park, CA 94025 USA \label{addr0}
\and
Stanford University, 450 Jane Stanford Way, Stanford, CA 94305 USA \label{addr1}
\and
Università degli Studi di Milano-Bicocca, Piazza dell'Ateneo Nuovo 1, 20126 Milano IT\label{addr2}
\and
Istituto Nazionale di Fisica Nucleare, Sezione di Milano-Bicocca, Piazza della Scienza 3, 20126 Milano IT \label{addr3}
\and 
Università degli Studi di Napoli Federico II, via Cinthia 21, I-80126 Napoli, IT\label{addr4}
\and 
Istituto Nazionale di Fisica Nucleare Sezione di Napoli, Complesso Universitario di Monte Sant'Angelo, Via Cintia, I-80126 Napoli IT\label{addr5}
\and 
Istituto Nazionale di Fisica Nucleare Sezione di Perugia, Via A. Pascoli, I-06123 Perugia IT\label{addr6}
}


\maketitle

\begin{abstract}
Dual-readout calorimeters achieve superior energy resolution by simultaneously measuring Cherenkov and scintillation signals for event-by-event electromagnetic fraction correction, making them attractive for next-generation Higgs factories. 
However, if a full waveform readout is required for time-based analysis to separate Cherenkov and scintillation signals, high off-detector data rates might present challenges. These challenges can be mitigated by real-time signal processing in front-end electronics.
We present a systematic comparison of machine learning (ML) and template fitting approaches for the separation of scintillation and Cherenkov light components in homogeneous dual-readout calorimeters across three representative crystal types.
ML models achieve comparable signal extraction performance at lower sampling rates than template fitting.
A single model trained over a range of incident particle energies demonstrates robust performance, and FPGA-compatible compression achieves latencies suitable for real-time application. 
This work establishes both baseline template fitting performance and ML-enhanced alternatives for crystal-based dual-readout calorimeters, offering practical pathways towards front-end feature extraction in future detector design.
\end{abstract}

\section{Introduction}
\label{sec:intro}

Dual-readout calorimeters~\cite{RevModPhys.90.025002,hirosky2024dualreadoutcalorimetryhomogeneouscrystals,refId0} represent a promising detector technology for the $ee$ stage of the Future Circular Collider (FCC-ee)~\cite{Abada2019,FCC:2025lpp}, where precise jet energy measurements are essential for physics drivers such as Standard Model measurements and Higgs boson studies. 
By simultaneously measuring both the scintillation (S) and Cherenkov (C) signals produced by particle showers, dual-readout calorimeters can achieve significantly improved hadronic energy resolution~\cite{Eno:2025ltc} through event-by-event determination of the electromagnetic shower fraction. In particular, a hybrid dual-readout calorimeter concept is recently proposed for the IDEA detector~\cite{Lucchini:2020bac,IDEAStudyGroup:2025gbt,Lucchini:2022vss}. This design consists of a homogeneous crystal electromagnetic section followed by a sampling calorimeter hadronic section. The implementation of dual-readout in the electromagnetic section requires simultaneous measurement of scintillation and Cherenkov photons produced within the same crystal volume. This can be achieved by exploiting both the wavelength and temporal features that distinguish scintillation light emission (typically slow and at a specific wavelength) from Cherenkov photons (promptly emitted over a continuum of wavelengths). 

Exploiting the temporal differences typically requires template fitting of the measured waveforms to accurately extract the C and S photon counts from the composite signal. This is a challenging task due to the vastly different temporal characteristics of the two components and the overlapping nature of their signals. Moreover, in future high-luminosity collider environments, the need to read out full waveforms at high event rates generates enormous data volumes, posing significant data rate challenges for the readout system.

For homogeneous crystal dual-readout calorimeters requiring waveform-based signal separation, conventional approaches employ template fitting of the measured waveforms, where a parametric model is optimized via iterative minimization. While effective at high sampling rates, this method is computationally expensive and incompatible with real-time requirements. Moreover, its performance degrades rapidly at reduced sampling rates, necessitating high-speed digitization with associated increases in power consumption, data bandwidth, and detector services.

A promising approach to address this signal processing challenge is edge machine learning (ML), where lightweight algorithms deployed directly in front-end readout electronics perform real-time feature extraction at the source, dramatically reducing off-detector data rates while maintaining low latency~\cite{gonski2026machinelearningheterogeneousedge}. 
Recent advances in embedded FPGA (eFPGA) technology~\cite{Gonski_2024} have demonstrated the feasibility of implementing real-time capable ML inference in reconfigurable ASICs (Technology Readiness Level 4~\cite{EC_H2020_WP_2016_2017}).
The concept of eFPGA-based front-end readout can enable ``smart" detectors for FCC-ee capable of intelligent and adaptable feature extraction, demonstrated for the use case of cluster counting in drift chambers~\cite{Yilmaz_2025_ML4PS}. 
However, the application of edge ML to continuous waveform decomposition in calorimetry presents distinct challenges. The task requires extracting scintillation and Cherenkov signal components from a composite waveform where these two components have vastly different temporal characteristics and overlap in time. This represents a fundamentally different problem from classification or single-feature regression, as the network must learn multivariate deconvolution in the presence of noise and limited sampling resolution.

In this work, we present the first systematic comparison of conventional template fitting to compact neural network models, in simulation of three representative scintillating crystals (BGO, BSO, and PWO), to motivate real-time dual-readout waveform analysis. The use of eFPGA technology to support this approach in on-detector readout thus provides a promising avenue to mitigate data rate challenges while enhancing performance of future dual readout front-end electronics.
\section{Samples}
\label{sec:samples}

The training and evaluation datasets are generated using Geant4~\cite{AGOSTINELLI2003250} Monte Carlo simulations of a hybrid dual-readout calorimeter consisting of a homogeneous crystal electromagnetic section, followed by a sampling hadronic calorimeter section, as described in~\cite{Lucchini:2020bac}. Electromagnetic and hadronic showers are simulated using electrons (e$^{-}$) and neutral kaons ($K^{0}_L)$, respectively. Each simulated event records the total number of Cherenkov photons produced in the ECAL module and the total deposited energy. The detected photoelectron counts are derived from these Monte Carlo truth values via light yield scaling and Poisson sampling:

\begin{align}
c &= \text{Poisson}(N_{\text{C}} \times \epsilon_C)\\
s &= \text{Poisson}(E_{\text{dep}} \times Y_S)
\end{align}

where $N_{\text{C}}$ is the total number of Cherenkov photons produced in the calorimeter, $E_{\text{dep}}$ is the deposited energy (GeV), $\epsilon_C = 0.09$ is the Cherenkov photon detection efficiency, and $Y_S$ is the crystal-specific scintillation light yield (phe/GeV). Each Monte Carlo event is resampled five times with independent Poisson realizations to augment the dataset.

For each sampled photon count pair $(c, s)$, a composite digitized waveform is constructed as follows. Photon arrival times are sampled from decay distributions that are instantaneous for Cherenkov and crystal-specific for scintillation, then convolved with the Single Photon Response (SPR) template obtained from laser calibration as described in~\cite{Alviggi:2026uzo}. SiPM dark counts randomly distributed in time at a rate of 10 MHz are also included in the waveform simulation. A random time shift $t_0$, drawn uniformly from the range [-2.5, 2.5] ns, is applied to both signal components to emulate timing variations. These variations might arise from physical effects such as different particle time of flight from the interaction point to the detector and the temporal development of electromagnetic and hadronic showers inside the calorimeter. The final waveform is the sum of the time-shifted Cherenkov and scintillation signals, dark counts, and a constant baseline offset. Each waveform comprises 6247 samples over a $[-400, 1599]$ ns window corresponding to a 3.125 GHz sampling rate, stored with ground truth labels $(c, s, t_0)$. These waveforms are subsequently downsampled to lower sampling rates for evaluation of ML model performance and resource efficiency.

\begin{figure}[tbh]
\centering
\includegraphics[width=0.5\textwidth]{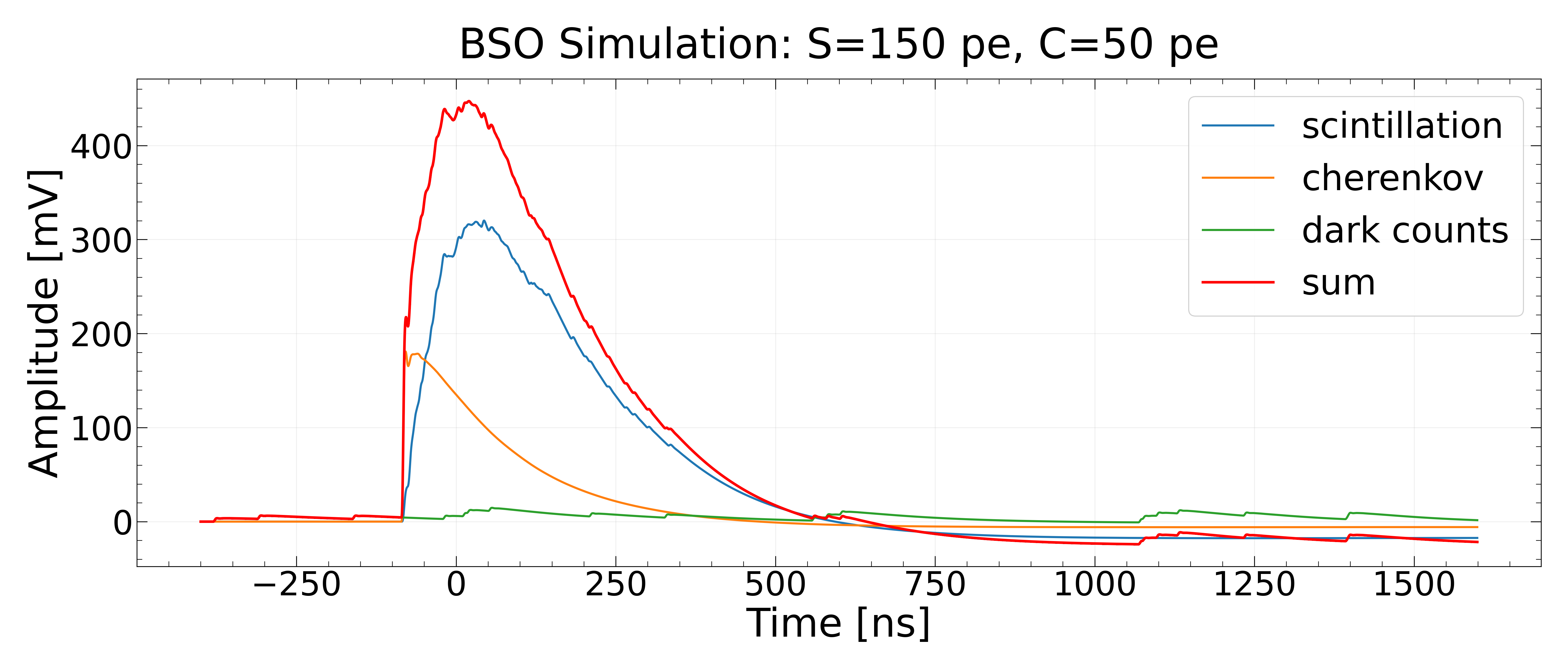}
\caption{\label{fig:wf_BSO_s150_c50} 
Representative composite waveform for BSO crystal with $50$ Cherenkov photons, $150$ scintillation photons, and $t_0=0$ ns. The individual contributions from scintillation, Cherenkov, and dark count components are shown separately.}
\end{figure}

\begin{table*}[hbtp]
\centering
\caption{\label{tab:crystal_properties} Properties of the three crystals studied in this work. Light yield values represent effective detected photons per unit energy 
after optical filtering~\cite{Alviggi:2026uzo}. The decay model describes the scintillation temporal response, with decay constants $\tau_1$ (fast) and $\tau_2$ (slow).}
\begin{tabular}{p{5.5cm}ccc}
\hline
Property & BGO & BSO & PWO \\
\hline
Scintillation light yield (phe/GeV) & 170 & 30 & 30 \\
Cherenkov light yield (phe/GeV) & 50 & 50 & 50 \\
Decay model & Bi-exp & Bi-exp & Single-exp \\
Fast decay constant $\tau_1$ (ns) & 50 & 22 & 7.5 \\
Slow decay constant $\tau_2$ (ns) & 320 & 98 & --- \\
\hline
\end{tabular}
\end{table*}

We evaluate signal extraction performance using three representative scintillating crystals with widely varying different characteristic scintillation decay times~\cite{Benaglia:2025pdf}: BGO, BSO, and PWO. These crystals span nearly two orders of magnitude in scintillation decay time and present distinct challenges for Cherenkov-scintillation decomposition. Cherenkov photons are emitted promptly (modeled as 
$\tau = 0.1$ ns exponential, effectively instantaneous), while 
scintillation follows crystal-specific decay distributions. The SPR templates used for waveform generation incorporate both 
the SiPM response to single photoelectrons and front-end electronics 
shaping effects, including low-frequency filtering characteristics. 
These effects were evaluated from laser calibration measurements~\cite{Alviggi:2026uzo} 
using 6$\times$6~mm$^2$ Hamamatsu S14160-6050HS SiPMs~\cite{Hamamatsu:S14160}.
A representative waveform generated for BSO crystal ($c=50$, $s=150$, $t_0=0$ ns) is shown in Figure~\ref{fig:wf_BSO_s150_c50}.

Table~\ref{tab:crystal_properties} summarizes the key properties of all three crystals. The light yield values correspond to the number of photoelectrons per unit energy detected on the SiPM dedicated to Cherenkov photon measurements. These values are based on laboratory and beam test measurements~\cite{Alviggi:2026uzo} and account for light collection effects, 
SiPM photodetection efficiency (PDE), and optical filtering to quench the scintillation signal. In particular, for BGO and BSO a low-pass SCHOTT UG11 filter is considered, while for PWO a high-pass Kodak 25 filter is assumed~\cite{Benaglia:2025pdf}. BGO has the highest scintillation light yield but the slowest decay, with a dominant 320 ns component that creates a long exponential tail. This results in a very large scintillation signal that overwhelms the Cherenkov component, making extraction of the small C signal particularly challenging. BSO represents an intermediate case with moderate light yield and a bi-exponential decay. 
PWO has the fastest scintillation response, producing signals that are nearly coincident with the prompt Cherenkov emission, given the slow SiPM response to single photoelectrons of about 50 ns. 
This strong temporal overlap makes PWO particularly challenging at reduced sampling rates, where temporal resolution is insufficient to separate the two components. In such cases, SiPMs with cell recharge times faster than 10 ns should be employed, or alternatively, the separation of Cherenkov and scintillation light should rely entirely on wavelength-based filtering as described in \cite{Benaglia:2025pdf}.

\begin{figure*}[htbp]
\centering
\includegraphics[width=0.95\textwidth]{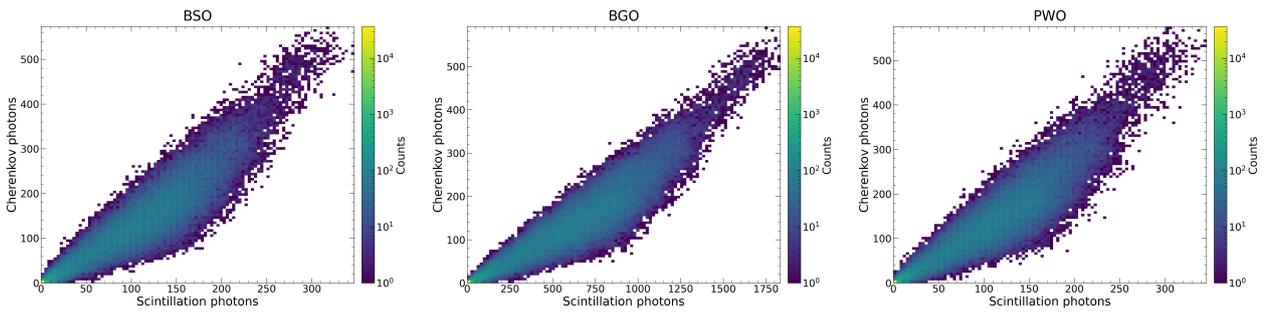}
\caption{\label{fig:cs_distributions}
Two-dimensional distributions of Cherenkov versus scintillation photon counts for 10 GeV K$^0_\text{L}$ events in BSO (left), BGO (center), and PWO (right) crystals. The color scale indicates event density.}
\end{figure*}

Figure~\ref{fig:cs_distributions} shows the distribution of C and S photon counts for 10 GeV K$^0_\text{L}$ events in the three crystals. BGO exhibits the widest range in $s$ due to its high light yield, while all three crystals show similar $c$ distributions since the Cherenkov yields after optical filtering are identical for all crystal candidates.

\section{Methodology}
\label{sec:methods}

To assess the feasibility of ML-based front end processing, the proposed ML method must satisfy two key criteria: meeting or exceeding performance for accurate determination of waveform ($c$, $s$, $t_0$), and synthesis for FPGA implementation to emulate the deployment on eFPGA ASICs. 
The ML method presented here is benchmarked against a conventional waveform template fitting approach.

\subsection{Template Fitting}

Template fitting serves as the baseline method for waveform decomposition. Normalized templates for Cherenkov and scintillation signals are pre-generated using high-statistics Monte Carlo simulation as follows. For each crystal, $10^7$ photon arrival times are sampled from the appropriate decay distribution, histogrammed with 0.32 ns resolution, convolved with the SPR template, and normalized to yield per-photon response functions \\ $C_{\text{template}}(t)$ and $S_{\text{template}}(t)$. These templates are converted to continuous interpolation functions to allow evaluation at arbitrary time shifts.

For each observed waveform, the three-parameter model
\begin{equation}
\text{WF(t)} = c \cdot C_{\text{template}}(t - t_0) + s \cdot S_{\text{template}}(t - t_0) + \text{offset}
\end{equation}
is fitted to the data using the \texttt{iminuit}~\cite{iminuit} optimizer with a two-stage minimization strategy. The baseline noise level is estimated from the standard deviation of the first 200 time samples (pre-trigger region) and used as a constant uncertainty across all bins. Physical constraints enforce $c, s \geq 0$ and bound the parameters to reasonable ranges based on expected photon counts and timing windows. Initial parameter values are estimated heuristically: $t_0$ from the half-maximum crossing time of the waveform rising edge, and $c$, $s$ both initialized to half the waveform peak amplitude, providing a label-free starting point robust across events. The fitting procedure employs a two-stage approach to ensure robust convergence: first, the Simplex algorithm~\cite{Dantzig1965} performs gradient-free global search to avoid local minima, followed by the MIGRAD~\cite{JAMES1975343} variable-metric algorithm for precise refinement and parameter uncertainty estimation.

While template fitting performs well at high sampling rates, it has several limitations. First, the iterative optimization requires massive function evaluations per waveform, making it computationally expensive and incompatible with real-time applications. Second, performance degrades rapidly at reduced sampling rates, as the waveform becomes under-constrained for three-parameter optimization.

\subsection{Deep Neural Network}

We develop a compact fully-connected neural network that directly regresses the signal parameters $(c, s, t_0)$ from the digitized waveform in a single forward pass. The network architecture is designed for FPGA deployability while maintaining competitive accuracy.
The model architecture is: Input $\to$ FC(16, ReLU) $\to$ FC(24, ReLU) $\to$ FC(8, ReLU) $\to$ FC(3, Linear), where the input dimension varies with 
the downsampling rate. 
The dataset is partitioned into training (70\%), validation (10\%), and test (20\%) sets using a fixed random seed for reproducibility. 
Both electron and kaon events contribute equally to ensure the training set covers a wide range of Cherenkov and scintillation 
photon counts representative of realistic detector conditions. With five samples per Monte Carlo event, the baseline dataset 
contains 140,000 training waveforms, 20,000 validation waveforms, and 40,000 test waveforms per energy point.

To evaluate performance at reduced data rates and enable compact deployment, the baseline 3.125 GHz waveforms are downsampled by integer factors $N \in \{10, 20, 50, 300\}$, yielding effective sampling rates of 312.5 MHz, 156.3 MHz, 62.5 MHz, and 10.4 MHz. Downsampling reduces power consumption and data bandwidth in the readout chain while simultaneously decreasing ML model complexity. This reduction directly lowers the first hidden layer parameter count, proportionally decreasing FPGA resource utilization and inference latency. Downsampled waveforms are normalized to $[0, 1]$ using a MinMaxScaler fitted on the training set. This preprocessing pipeline is deterministic and implementable in fixed-point arithmetic on FPGA.

The three regression targets are scaled to comparable numerical ranges via crystal-specific normalization constants:
$$\mathbf{y}_{\text{scaled}} = \left[\frac{c}{c_{\text{norm}}}, \frac{s}{s_{\text{norm}}}, \frac{t_0}{t_{0,\text{norm}}}\right]$$
where $(c_{\text{norm}}, s_{\text{norm}}, t_{0,\text{norm}}) = (500, Y_S/30 \times 220, 2.5)$ for crystal-dependent scintillation yield $Y_S$. This ensures all output labels occupy similar ranges during training. At inference, predictions are inverse-transformed to physical units. The network is trained using the Adam optimizer~\cite{kingma2017adammethodstochasticoptimization} with a weighted mean squared error loss:
\begin{equation}
    \mathcal{L} = \frac{1}{N}\sum_{i=1}^N \mathbf{w}^\top (\hat{\mathbf{y}}_i - \mathbf{y}_i)^2
\end{equation}

where $\hat{\mathbf{y}} = (\hat{c}, \hat{s}, \hat{t}_0)$ and weights $\mathbf{w} = (4.0, 1.0, 0.3)^\top$ reflect the relative importance of each output. The learning rate begins with a 20-epoch linear warmup from $10^{-5}$ to $0.01$, then reduces adaptively with factor 0.7 and patience 40 epochs. Training uses batch size 512 with early stopping at patience 30 epochs.

To ensure successful convergence, a failure detection mechanism monitors validation loss at epoch 40: if $\mathcal{L}_{\text{val}} > 0.08$, training is aborted and restarted with a different random seed. This addresses the sensitivity of small networks to weight initialization. Trained models are synthesized to FPGA firmware using \hlsfml~\cite{fastml_hls4ml,Duarte:2018ite}, an open-source tool that converts Keras/TensorFlow models to High-Level Synthesis C++ code and generates hardware implementations. For production deployment, a model compression pipeline incorporating magnitude-based pruning and quantization-aware training (QAT) using QKeras~\cite{Coelho:2020zfu} is applied to reduce resource consumption while maintaining accuracy.

\section{Results}
\label{sec:results}

\subsection{Performance}

Performance is evaluated using the 68th percentile error (err68) as the primary metric, corresponding to approximately $1\sigma$ for Gaussian-distributed errors. For Cherenkov and scintillation photon counts, we report the 68th percentile relative error, defined as:
\begin{equation}
    \text{err68} = P_{68}\left(\frac{|\hat{q} - q|}{q} \times 100\%\right)
    \label{eq:err68}
\end{equation}
where $\hat{q}$ is the predicted value and $q$ is the true value. For timing extraction, we report the timing resolution (68th percentile absolute error in nanoseconds), since many events have $t_0 \approx 0$, making relative errors ill-defined.
Since relative errors are undefined when the true photon count is zero, events with $c=0$ or $s =0$ are excluded from evaluation.

\begin{figure}[htbp]
\centering
\includegraphics[width=0.5\textwidth]{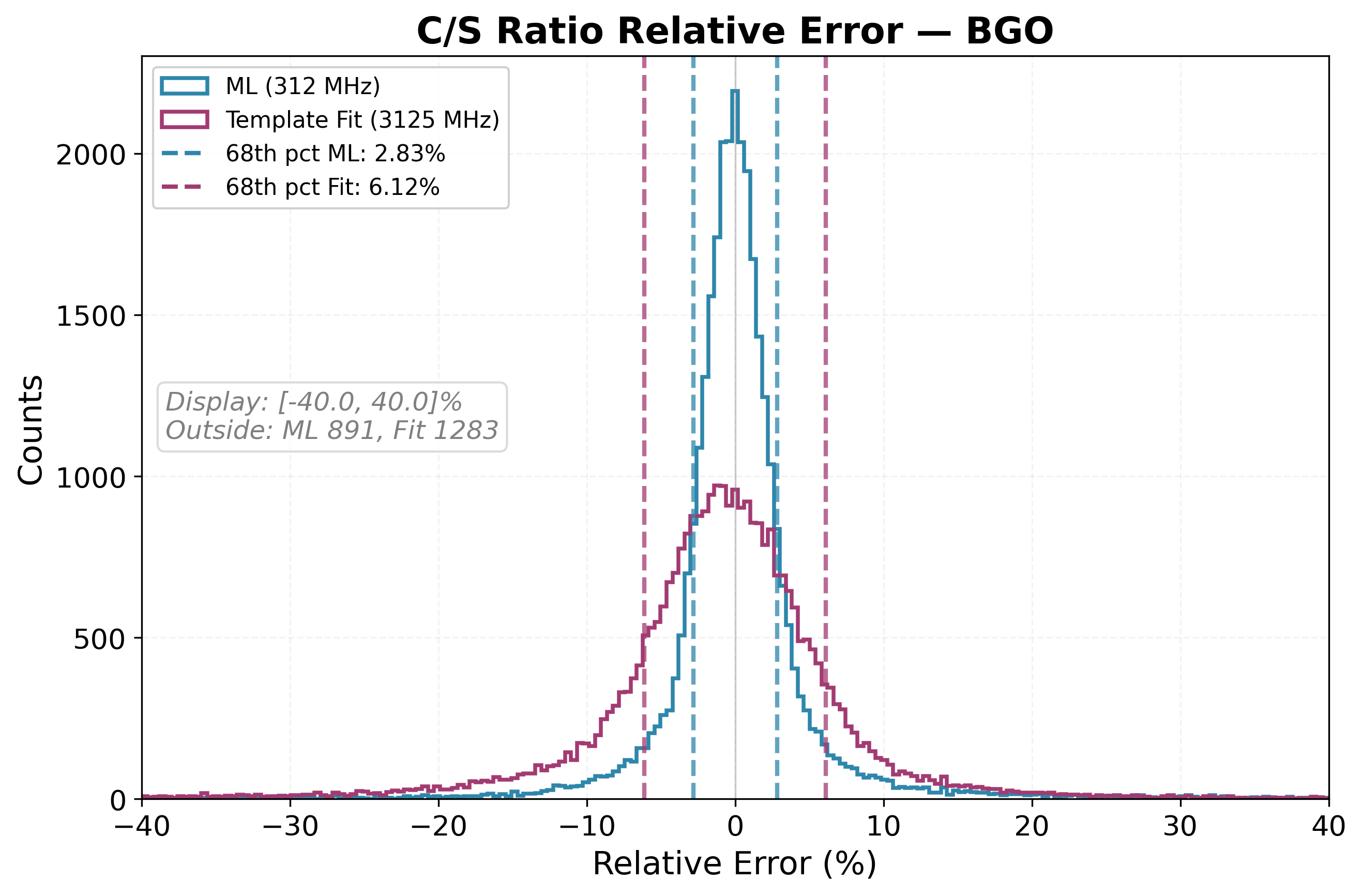}
\caption{\label{fig:cs_ratio_dist}
Distribution of C/S ratio relative error for BGO crystal, comparing the ML approach at 312.5 MHz with template fitting at 3.125 GHz (baseline sampling rate). Vertical dashed lines mark the 68th percentile error for each method.}
\end{figure}

Figure~\ref{fig:cs_ratio_dist} illustrates the definition of the 68th percentile metric using C/S ratio relative error distributions for BGO crystal. The ML approach operating at 312.5 MHz achieves $\text{err68} = \text{2.83}\%$, significantly outperforming template fitting at the full 3.125 GHz sampling rate. This example illustrates the potential of the ML approach to achieve competitive or superior accuracy at reduced sampling rates, motivating the systematic performance comparisons across crystals and operating conditions presented below.

\subsubsection{Sampling Rate Dependence}\label{sampling-study}

\begin{figure*}[htbp]
\centering
\includegraphics[width=0.95\textwidth]{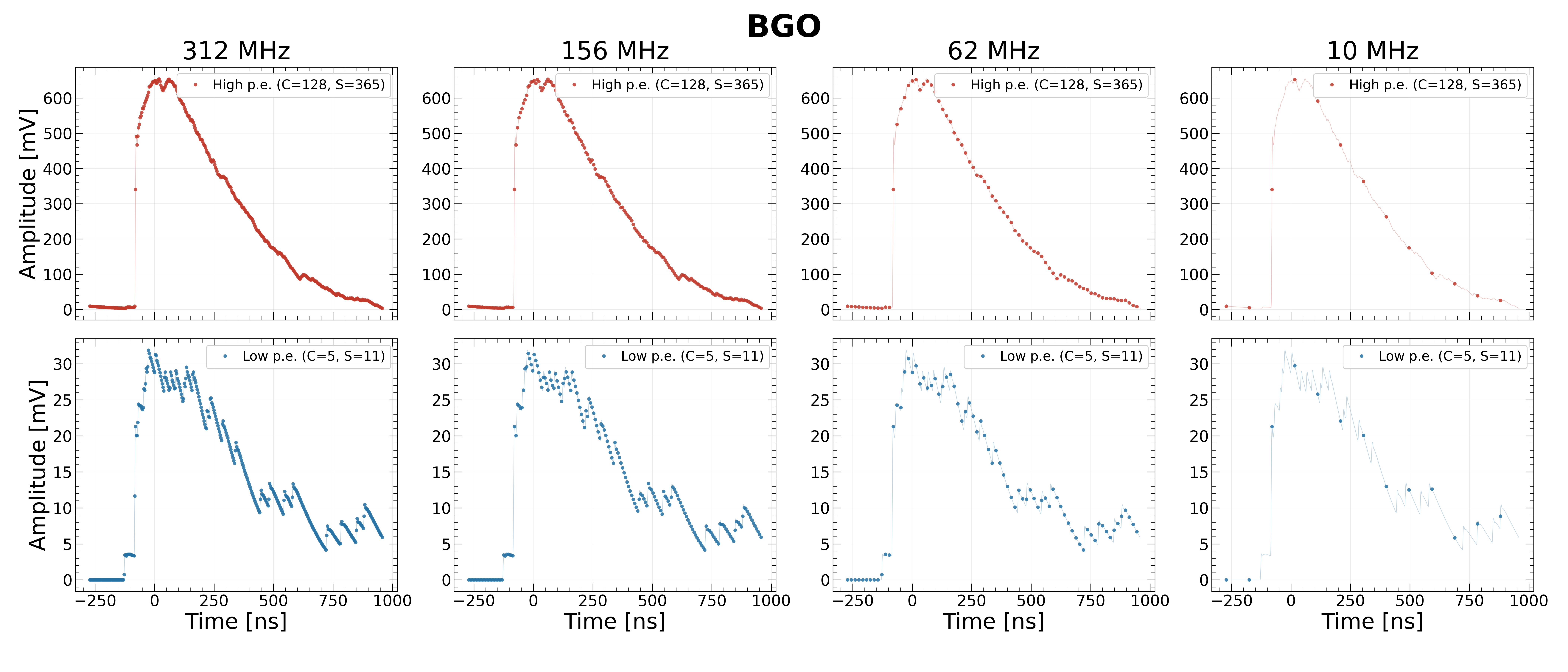}
\includegraphics[width=0.95\textwidth]{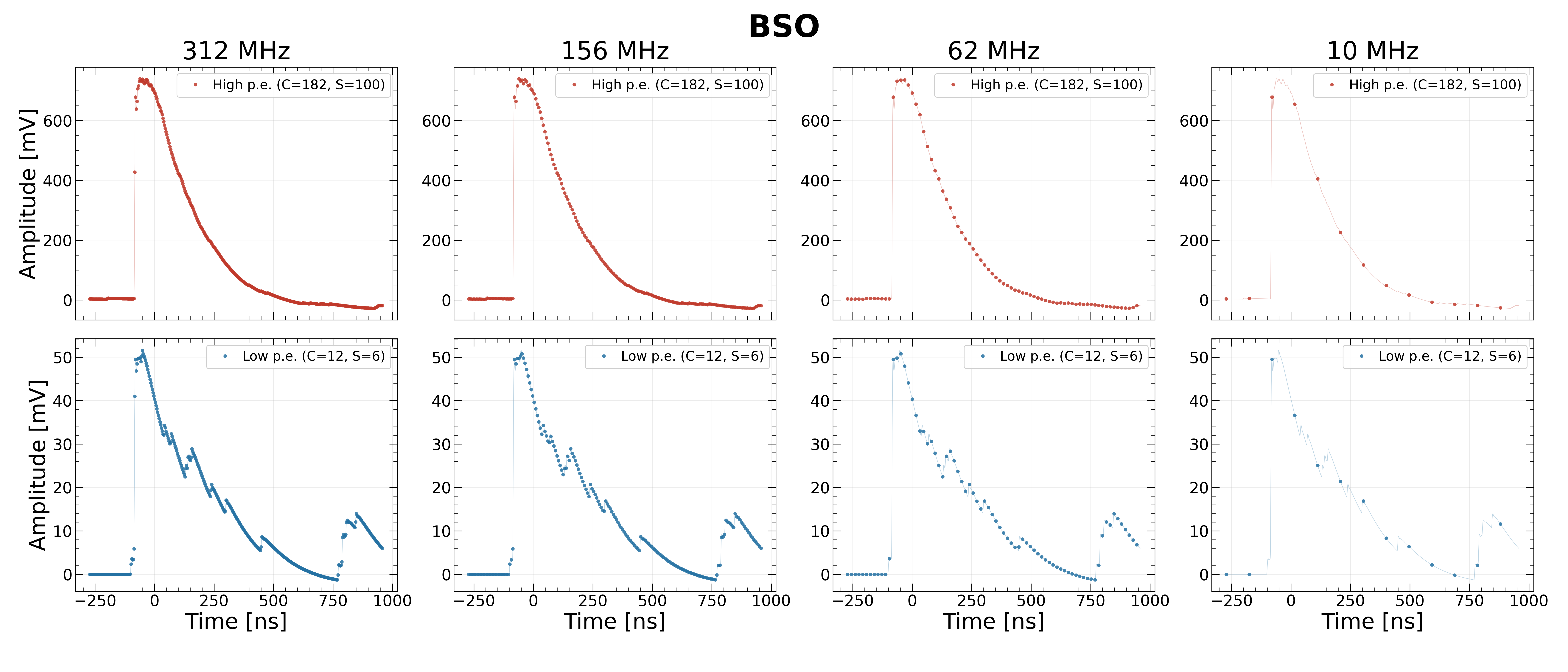}
\includegraphics[width=0.95\textwidth]{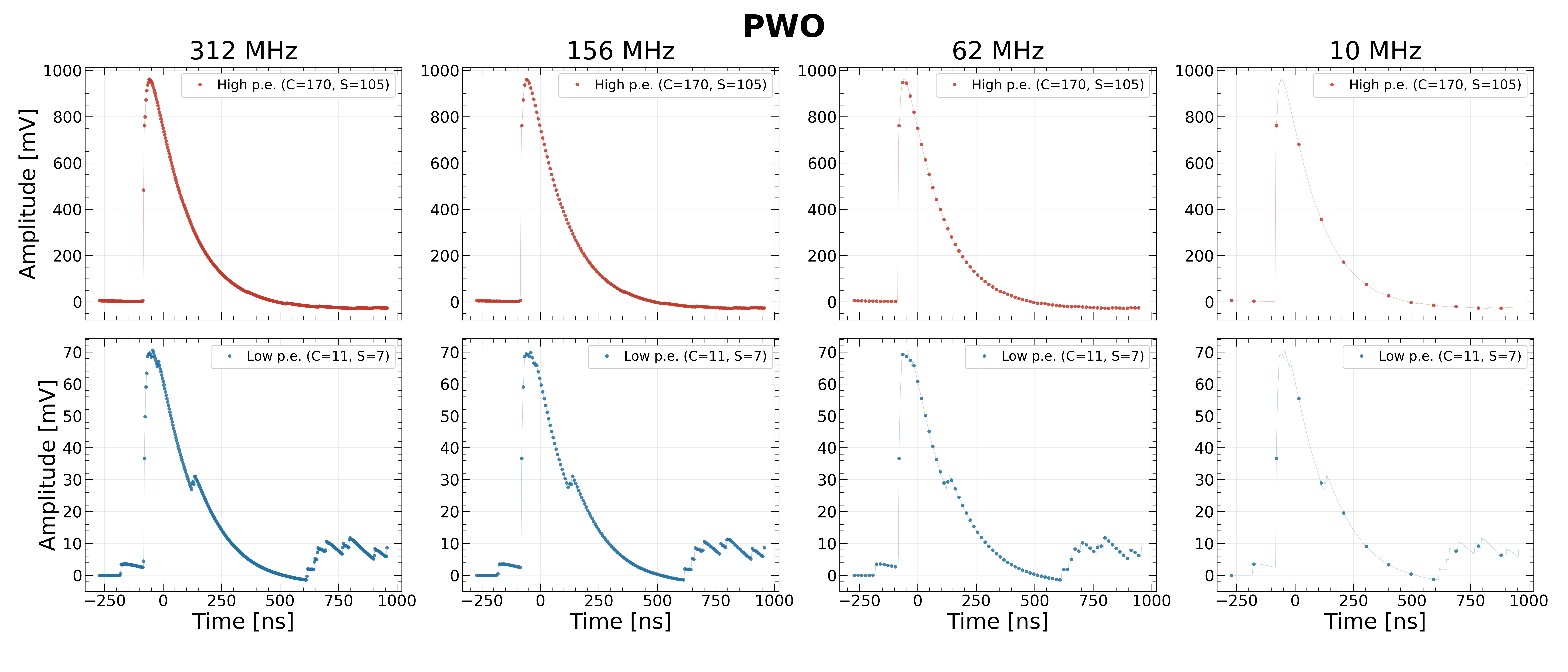}
\caption{\label{fig:waveform_sampling}
Digitized waveforms at different sampling rates for BGO (top), 
BSO (middle), and PWO (bottom) crystals. For each crystal, 
representative high photoelectron (top row, red) and low 
photoelectron (bottom row, blue) events are shown at 312.5, 
156.3, 62.5, and 10.4 MHz sampling rates. Sampled points are 
shown as circles, with the original 3.125 GHz waveform displayed 
as a faint background.}
\end{figure*}

We evaluate performance across sampling rates from 312.5 MHz to 10.4 MHz for all three crystals at fixed 10 GeV beam energy. As described in Section 3.2, these rates correspond to 10×, 20×, 50×, and 300× downsampling of the baseline 3.125 GHz waveforms, with input dimensionalities ranging from approximately 600 to 15 features. Template fitting is also further evaluated at 3.125 GHz to establish baseline performance at full sampling rates. The impact of reduced sampling rates on waveform quality is 
illustrated in Figure~\ref{fig:waveform_sampling} for all three 
crystals at both high and low photoelectron counts. At 10.4 MHz, only 15 sample points capture the entire signal, making feature extraction challenging, especially for low photoelectron events where SiPM dark count noise becomes comparable to signal amplitude.

\begin{figure*}[htbp]
\centering
\includegraphics[width=0.9\textwidth]{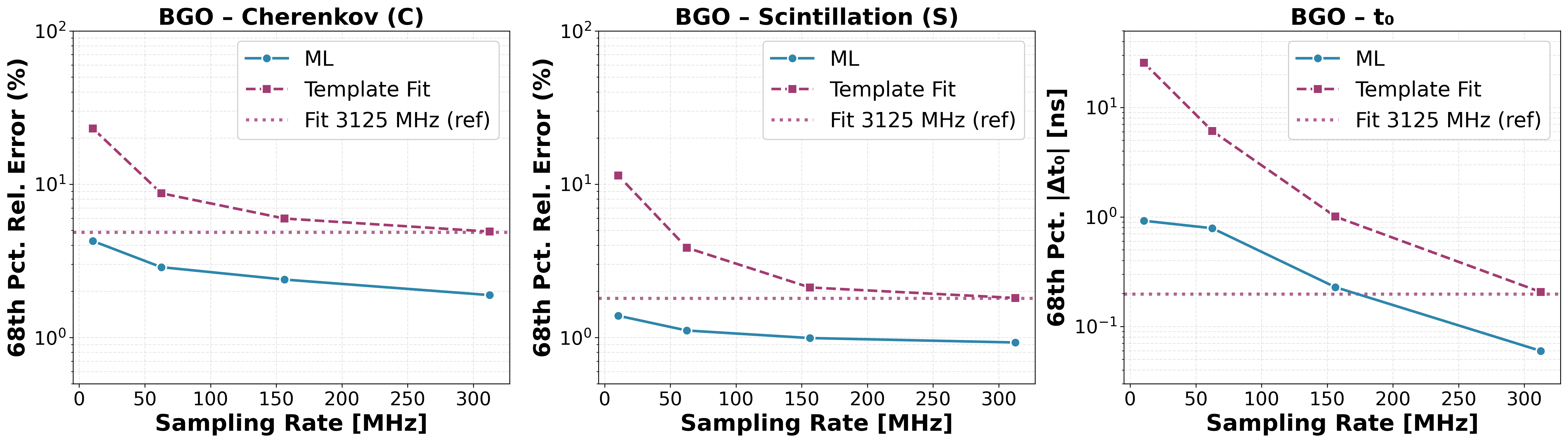}
\includegraphics[width=0.9\textwidth]{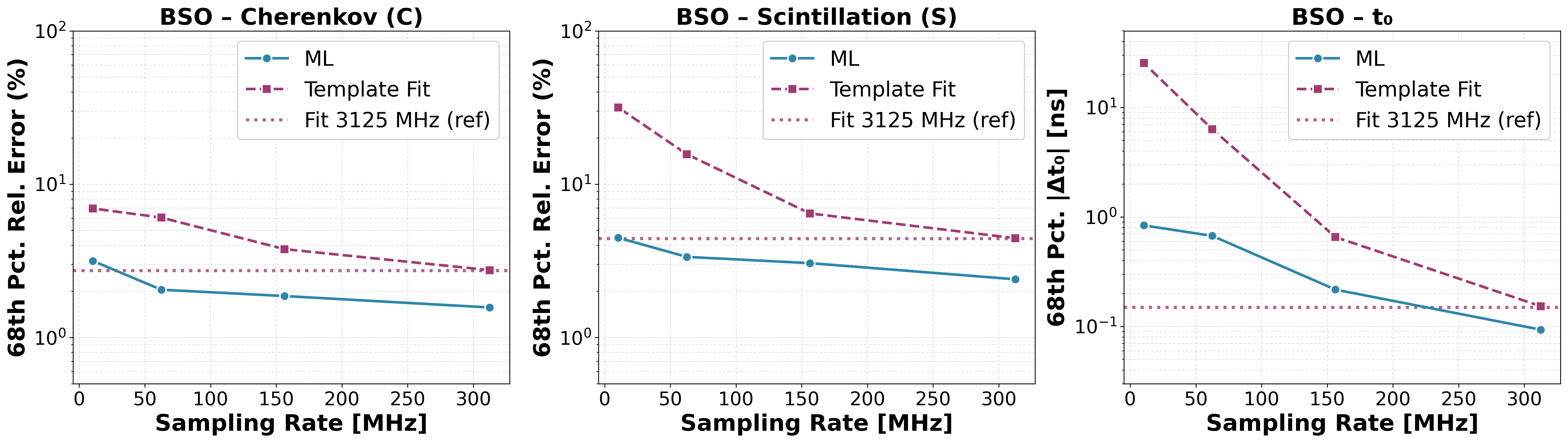}
\includegraphics[width=0.9\textwidth]{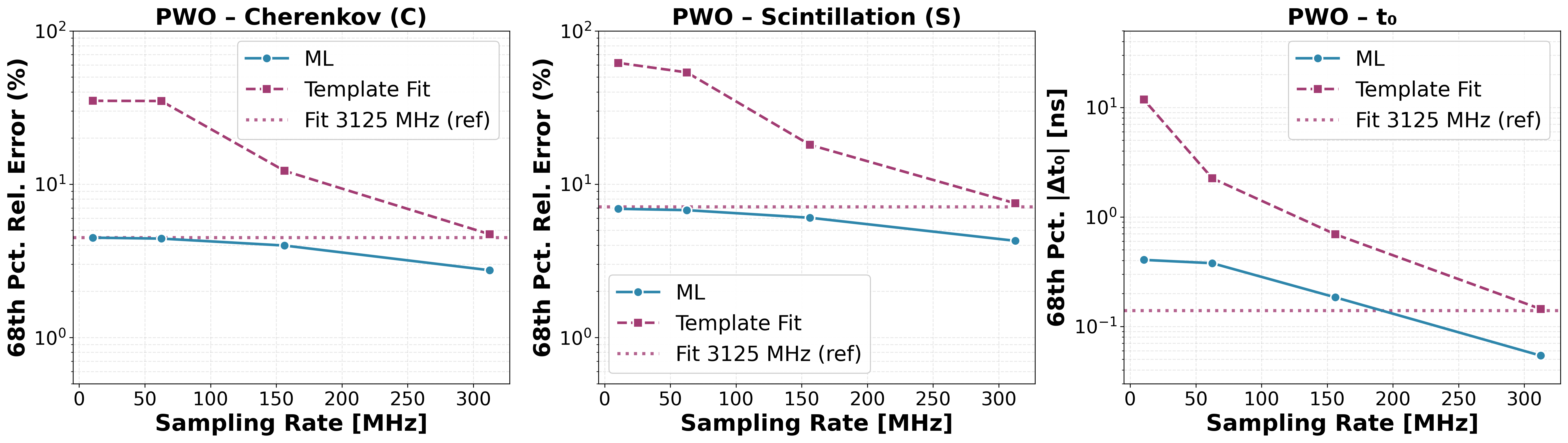}
\caption{\label{fig:results_vs_rate}
Performance versus sampling rate for BGO (top), BSO (middle), and PWO (bottom) crystals at 10 GeV. For each crystal, the 68th percentile error is shown for Cherenkov photon count (left), scintillation photon count (center), and arrival time $t_0$ (right). ML results (blue solid lines) are compared with template fitting at 312.5 MHz (purple solid lines) and at 3.125 GHz baseline (purple horizontal dashed line, shown as reference).}
\end{figure*}

Figure~\ref{fig:results_vs_rate} shows the 68th percentile error for C, S, and $t_0$ extraction as a function of sampling rate.
At the lowest sampling rate (10.4 MHz), the ML approach achieves highly competitive performance compared to template fitting at full 3.125 GHz sampling rate, particularly for C and S extraction. For BGO and BSO, ML at 10.4 MHz yields $\text{err68}_{\text{C}}$ and $\text{err68}_{\text{S}}$ within $5\%$. PWO, which presents the greatest challenge due to temporal overlap of C and S signals, shows similar trends, with ML at reduced rates remaining competitive with template fitting at substantially higher sampling rates. 

For timing extraction, ML at low sampling rates does not surpass template fitting at full sampling rate.
This is attributable to the sampling interval increasing from 0.32 ns to 96 ns, exceeding the timing jitter range and causing fundamental information loss beyond any algorithm's ability to recover. 
This represents an inherent physical constraint of undersampling rather than a limitation of the ML method. 
Nonetheless, ML achieves better $t_0$ extraction than template fitting when both operate at the same reduced rate, demonstrating more efficient use of available temporal information.

At higher sampling rates, the performance gap between ML and template fitting narrows for C and S extraction, though ML maintains a consistent advantage. For timing extraction, the difference between methods becomes less pronounced at high sampling rates, reflecting that both approaches can leverage the improved temporal resolution. 
The template fitting method at the full 3.125 GHz provides only marginal improvement over 312.5 MHz.

\subsubsection{Incident Energy Dependence}

To ensure the model performance generalizes well across particle energy, ML models are trained using combined datasets from multiple beam energies (1, 5, 10, and 30~GeV) to produce a single universal model capable of operating across the full energy range. This multi-energy training strategy decouples the network from energy-specific features and enables robust performance at arbitrary intermediate energies. 

\begin{figure*}[tbh]
\centering
\includegraphics[width=0.9\textwidth]{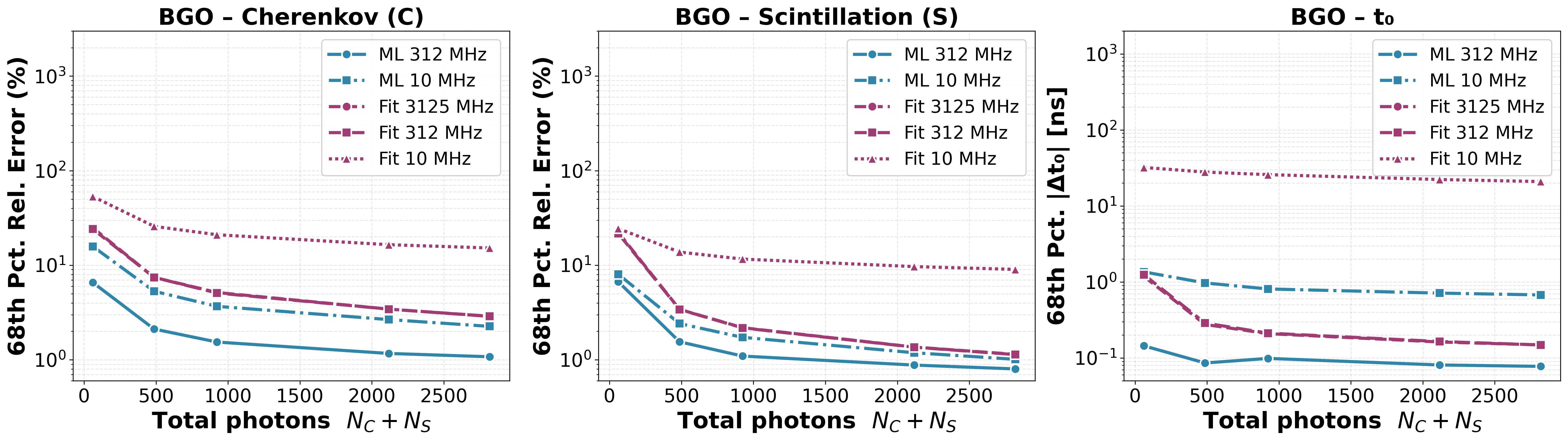}
\includegraphics[width=0.9\textwidth]{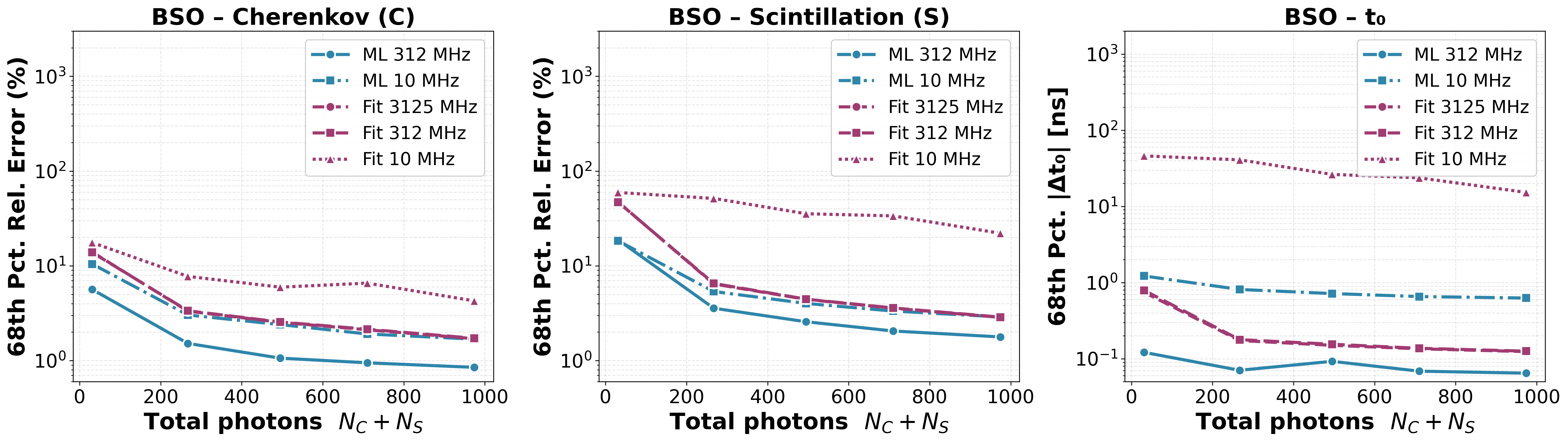}
\includegraphics[width=0.9\textwidth]{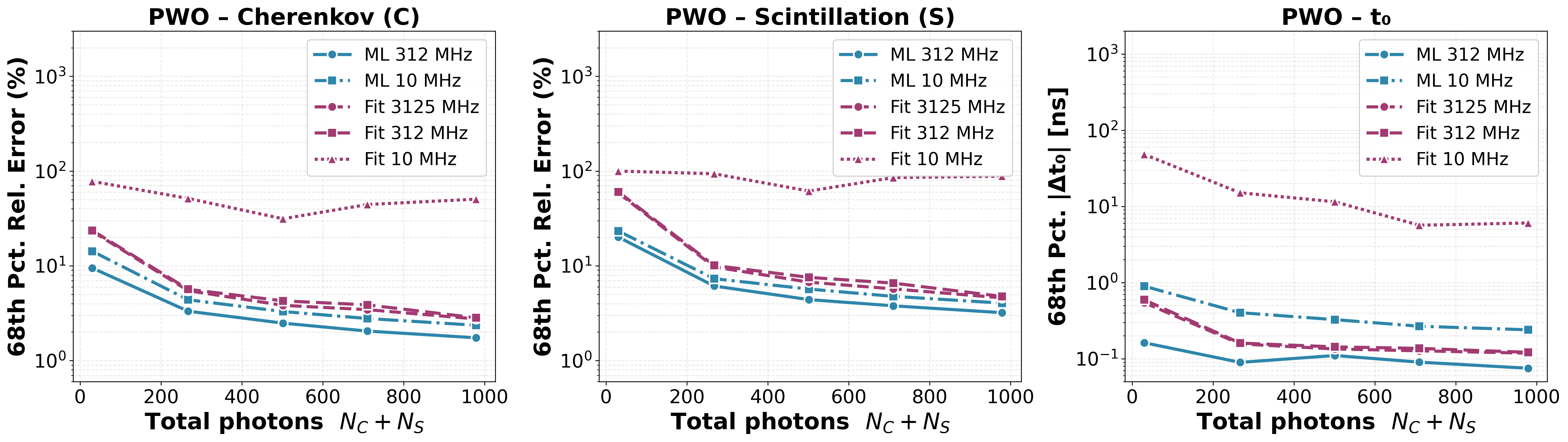}
\caption{\label{fig:results_vs_e}
Performance versus total detected photon count for BGO (top), BSO (middle), and PWO (bottom) crystals. The 68th percentile error is shown for Cherenkov photon count (left), scintillation photon count (center), and arrival time $t_0$ (right). ML results (blue lines) are shown at two sampling rates (312.5 MHz and 10.4 MHz), using models trained on combined 1, 5, 10, and 30~GeV datasets. Template fitting results (purple lines) are shown at three sampling rates (3.125 GHz, 312.5 MHz, and 10.4 MHz). Both methods are evaluated on identical held-out 20\% test sets.}
\end{figure*}

In contrast to the sampling rate studies in~\ref{sampling-study}, which generate waveforms from the total energy deposited across the entire ECAL, the evaluation presented here operates at the individual readout channel level -- each waveform is generated from the energy deposited in a single calorimeter channel (the highest-energy channel per event). This per-channel approach better reflects realistic detector conditions, where each SiPM channel reads out only a fraction of the total shower energy, resulting in lower signal amplitudes. For each crystal, the dataset consists of 30,000 simulated events per particle type at each of four beam energies. The ML model is trained on 80\% of these events (168,000 training + 24,000 validation across all energies combined), and evaluated on the remaining held-out 20\% test set, yielding 48,000 total evaluation events per crystal. Template fitting is evaluated on the same held-out test set to ensure a fair comparison.             

Figure~\ref{fig:results_vs_e} compares the performance of the multi-energy ML model with template fitting as a function of total detected photon count ($N_C + N_S$). Events from all four beam energies are combined and binned by total photon count, providing a unified view of reconstruction performance across the full signal range.
At 312.5 MHz, the ML approach demonstrates excellent performance across the full photon count range for all three crystals.
At 10.4 MHz, the ML performance exhibits minor degradation. This generalization capability is particularly valuable for applications where beam energy varies or is not precisely known a priori.
Timing resolution is generally stable across photon count bins at fixed sampling rate. Unlike photon count extraction, timing performance shows minimal crystal dependence, as $t_0$ reflects the shower initiation time rather than scintillation decay dynamics. At 312.5 MHz, ML achieves timing resolutions of 0.10--0.14 ns across all three crystals, compared to template fitting at 3.125 GHz (0.39--0.66 ns). At 10.4 MHz sampling rate, ML maintains resolutions of 0.74--1.17 ns, while template fitting degrades significantly to 29--45 ns. As with the sampling rate studies, we find that the ML method meets or exceeds performance of the high-sampling template fitting method while using an order of magnitude fewer input waveform samples.

\subsection{Resource Evaluation}
\label{subsec:fpga}

\begin{figure}[htbp]
\centering
\includegraphics[width=0.5\textwidth]{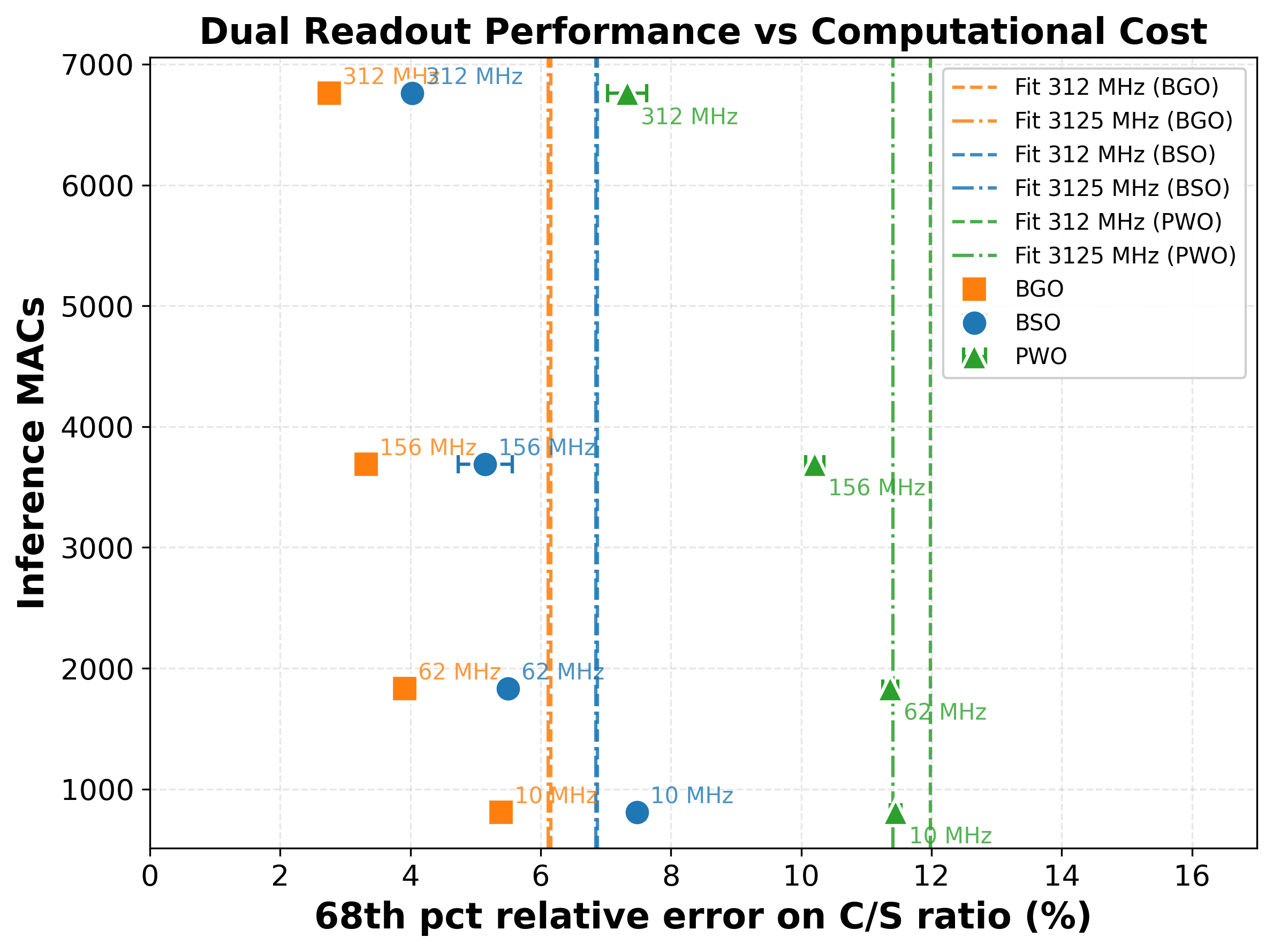}
\caption{\label{fig:mae_cs_ratio}
Computational cost versus C/S ratio extraction performance for uncompressed ML models across three crystals and four sampling rates. Vertical dashed lines indicate template fitting performance at 312.5 MHz (dash) and 3.125 GHz (dash-dot) for each crystal.}
\end{figure}

\begin{table*}[htbp]
\centering
\caption{FPGA resource utilization and performance for compressed ML models synthesized via \hlsfml. Models are optimized using pruning and quantization-aware training (QAT). For comparison, template fitting results at 3.125 GHz are shown. Bold values indicate cases where ML outperforms template fitting.}
\label{tab:fpga_resources}
\small
\begin{tabular}{l|lccccccc|c}
\hline
Crystal & Sampling & Compression & Est. Freq & Latency & LUT & FF & DSP & ML & Fit \\
 & Rate & Strategy & [MHz] & [ns] & & & & err68 & err68 \\
 & & & & & & & & & @3GHz \\ 
\hline
BGO & 10 MHz & 40\% sparsity, $\langle 10,1 \rangle$ QAT & 57.5 & 25 & 36,879 & 213 & 0 & \bf{5.84\%} & 6.12\% \\
 & 312 MHz & 30\% sparsity, $\langle 10,2 \rangle$ QAT & 46.7 & 25 & 312,914 & 3,933 & 0 & \bf{3.08\%} & 6.12\% \\
\hline
BSO & 10 MHz & 50\% sparsity, $\langle 10,2 \rangle$ QAT & 57.7 & 25 & 30,471 & 213 & 0 & 8.14\% & 6.85\% \\
 & 312 MHz & 30\% sparsity, $\langle 10,4 \rangle$ QAT & 46.6 & 25 & 328,953 & 3,933 & 0 & \bf{4.18\%} & 6.85\% \\
\hline
PWO & 10 MHz & 50\% sparsity, $\langle 10,2 \rangle$ QAT & 57.7 & 25 & 30,321 & 213 & 0 & 11.47\% & 11.41\% \\
 & 312 MHz & 30\% sparsity, $\langle 10,4 \rangle$ QAT & 46.6 & 25 & 311,282 & 3,933 & 0 & \bf{7.40\%} & 11.41\% \\
\hline
\multicolumn{10}{l}{\footnotesize Crystal parameters (light yield in phe/GeV, decay constants in ns):} \\
\multicolumn{10}{l}{\footnotesize BGO: $c$=50, $s$=170, $\tau_1$=50, $\tau_2$=320; BSO: $c$=50, $s$=30, $\tau_1$=22, $\tau_2$=98; PWO: $c$=50, $s$=30, $\tau$=7.5.} \\
\end{tabular}
\end{table*}

Before applying model compression, we first examine the trade-off between computational cost and physics performance for the baseline ML models. Figure~\ref{fig:mae_cs_ratio} shows the inference cost, measured in multiply-accumulate operations (MACs), versus the 68th percentile error on C/S ratio extraction for all three crystals across different sampling rates. MACs quantify inference cost as the sum of 
$d_{\text{in}} \times d_{\text{out}}$ over all dense layers. Since input 
dimension decreases with downsampling, MACs scale proportionally with sampling rate. We focus on C/S ratio error as the primary performance metric in this section because it directly relates to the dual-readout calorimeter's core capability.

The results reveal a clear Pareto frontier: for each crystal, models trained at different sampling rates span a range of computational costs with corresponding performance levels. Notably, all ML models, regardless of sampling rate, achieve C/S ratio errors competitive with or superior to template fitting (vertical dashed lines). This Pareto analysis demonstrates that substantial computational savings are possible by selecting appropriate sampling rates based on the target performance requirements. To further reduce FPGA resource consumption for deployment, we apply magnitude-based pruning and quantization-aware training (QAT) to these baseline models. Pruning levels and quantization bit-widths are optimized per crystal and sampling rate to minimize resource utilization while maintaining err68$_{\text{C/S}}$ close to the uncompressed baseline.

Table~\ref{tab:fpga_resources} summarizes the FPGA resource requirements and performance for the compressed models. The quantization format $\langle \text{total bits}, \text{integer bits} \rangle$ indicates fixed-point precision. All models achieve inference latencies below 25 ns at estimated clock frequencies of 46-58 MHz, which are considered to be compatible with potential FCC-ee bunch crossing rates. 
Resource utilization scales with sampling rate: models operating at 10.4 MHz require approximately 30k LUTs and 213 FFs, while 312.5 MHz models use approximately 300--330k LUTs and approximately 4k FFs. 
No DSP blocks are required due to the aggressive quantization, which converts all multiplications to shift-and-add operations. 
The elimination of DSPs is particularly valuable in \\ resource-constrained environments where DSPs are shared among multiple signal processing tasks.
Performance benchmarks are also provided for context. 
ML models at 10.4 MHz achieve competitive or superior $\text{err68}_{\text{C/S}}$ compared to template fitting at 3.125 GHz, while requiring orders of magnitude lower computational resources and data throughput. This demonstrates the viability of ML-based dual-readout signal extraction for real-time front-end processing in future high-luminosity colliders.
\section{Conclusions}

We have demonstrated that compact neural networks can effectively extract Cherenkov and scintillation signal components and timing information from dual-readout calorimeter waveforms with competitive performance compared to conventional template fitting at substantially reduced sampling rates. 
ML models at 312.5 MHz consistently outperform template fitting at the same rate and frequently match template fitting at 3.125 GHz baseline, enabling proportional reductions in ADC power consumption and data bandwidth. 
A single ML model trained on 1-30 GeV datasets generalizes robustly across energies without retraining. 
Model compression via pruning and quantization-aware training yields FPGA implementations with $\leq25$ ns latency and 30k-330k LUTs, demonstrating real-time deployment feasibility. 
Future work will focus on test beam validation and hardware characterization to demonstrate end-to-end benefits for eFPGA-based readout. 
This work establishes front-end ML as a viable tool for high-performance dual-readout calorimetry readout within the context of future FCC-ee detectors.

\begin{acknowledgements}
This work received funding from the U.S. Department of Energy under contract number DE-AC02-76SF00515 and the European Union in the framework of the Next Generation EU program, Mission 4, Component 1, CUP H53D23001120006.
\end{acknowledgements}



\bibliographystyle{unsrt}
\bibliography{main.bib}

\end{document}